\documentstyle[twoside,fleqn,espcrc2]{article}
\newcommand{\eqn}[1]{(\ref{#1})}
\def\bfone{\relax{\rm 1\kern-.35em 1}}
\def\bfzero{\relax{\rm I\kern-.18em 0}}
\def\IG{\relax\,\hbox{$\inbar\kern-.3em{\rm G}$}}
\font\cmss=cmss10 \font\cmsss=cmss10 at 7pt

\def\inbar{\vrule height1.5ex width.4pt depth0pt}
\def\IC{\relax\,\hbox{$\inbar\kern-.3em{\rm C}$}}
\def\IG{\relax\,\hbox{$\inbar\kern-.3em{\rm G}$}}
\def\IB{\relax{\rm I\kern-.18em B}}
\def\ID{\relax{\rm I\kern-.18em D}}
\def\IL{\relax{\rm I\kern-.18em L}}
\def\IF{\relax{\rm I\kern-.18em F}}
\def\IH{\relax{\rm I\kern-.18em H}}
\def\II{\relax{\rm I\kern-.17em I}}
\def\IN{\relax{\rm I\kern-.18em N}}
\def\IP{\relax{\rm I\kern-.18em P}}
\def\IQ{\relax\,\hbox{$\inbar\kern-.3em{\rm Q}$}}
\def\bfzero{\relax\,\hbox{$\inbar\kern-.3em{\rm 0}$}}
\def\IR{\relax{\rm I\kern-.18em R}}
\def\ZZ{\relax\ifmmode\mathchoice
{\hbox{\cmss Z\kern-.4em Z}}{\hbox{\cmss Z\kern-.4em Z}}
{\lower.9pt\hbox{\cmsss Z\kern-.4em Z}}
{\lower1.2pt\hbox{\cmsss Z\kern-.4em Z}}\else{\cmss Z\kern-.4em
Z}\fi}
\def\IE{\relax{{\rm I\kern-.18em E}}}
\def\IA{\relax{\hbox{{\rm A}\kern-.82em {\rm A}}}}
\newcommand{\AmS}{{\protect\the\textfont2
  A\kern-.1667em\lower.5ex\hbox{M}\kern-.125emS}}
\title{ Supersymmetry and First Order Equations for Extremal States:\\
Monopoles, Hyperinstantons, Black--Holes and $p$--branes}
\author{Pietro
Fr\'e\address{Dipartimento di Fisica Teorica, Universit\`a di Torino,
\\ Via P. Giuria 1, I-10125 TORINO, Italy\\
Email: Fre@to.infn.it }
\thanks{Research supported in part by   EEC  under TMR contract
 ERBFMRX-CT96-0045 } }
\begin{document}
\begin{abstract}
In this lecture I review recent results on the first order equations
describing BPS extremal states, in particular $N=2$ extremal
black--holes. The role of special geometry is emphasized also in the
rigid theory and a comparison is drawn with the supersymmetric derivation of
instantons and hyperinstantons in topological field theories. Work in
progress on the application of solvable Lie algebras to the
discussion of BPS states in maximally extended supergravities is
outlined.
\end{abstract}
\maketitle
\section{INTRODUCTION}
Supersymmetry has been the most powerful tool to advance
our understanding of quantum field theory and, in its $N=1,D=4$ spontaneously broken
form, it might also be experimentally observable. The locally $N$--extended
supersymmetric field theories, namely $N$--extended supergravities,
are interpreted as the low energy effective actions of various superstring
theories in diverse dimensions. However, since the duality revolution of
two years ago \cite{sumschwarz}, we know that all superstring models
(with their related effective actions) are just different corners of
a single non--perturbative quantum theory that includes, besides
strings also other $p$--brane excitations. Indeed in dimensions $D
\ne 4$ the duality rotations \cite{mylec} from
electrically to magnetically charged particles (= $0$--branes) generalize  to
transformations exchanging the perturbative elementary $p$--brane excitations
of a theory with the non--perturbative solitonic
$D-p-4$--brane excitations of the same theory. The mass per unit world--volume of
these objects is lower bounded by the value of the topological central
charge according to a generalization of the classical Bogomolny bound
on the monopole mass. In the recent literature,
the states saturating this lower bound are named  BPS saturated states
\cite{BPS} and play a prominent role in establishing the exact duality
symmetry of the quantum theory since they are the lowest lying stable
states of the non perturbative spectrum. They are characterized by the
fact that they preserve, in modern parlance, $1/2$ of the original
supersymmetries. What this actually means is that there is a suitable
projection operator $\IP^2_{BPS} =\IP_{BPS}$ acting on the supersymmetry charge
$Q_{SUSY}$, such that:
\begin{equation}
 \left(\IP_{BPS} \,Q_{SUSY} \right) \, \vert \, \mbox{BPS state} \, >
 \,=  \, 0
 \label{bstato}
\end{equation}
Since the supersymmetry transformation rules of any supersymmetric
field theory are linear in the first derivatives of the fields
eq.\eqn{bstato} is actually a {\it system of first order differential
equations} such that any solution of \eqn{bstato} is also a solution
of the {\it second order field equations} derived from the action
but not viceversa. Hence the case of BPS states, that includes
extremal monopole and black--hole configurations, is an instance of
the relation existing between {\it supersymmetry and first--order square
roots of the classical field equations}. Another important example of this
relation is provided by the {\it the topological twist}
\cite{topftwist_1,topftwist_2,topf4d_8} of
supersymmetric theories to {\it topological field theories}
\cite{wittft}. What happens here is that, after Wick rotation to the
Euclidean region, there is another projection operator $\IP_{BRST}^2=
\IP_{BRST}$ acting on the supersymmetry charge $Q_{SUSY}$, such that:
\begin{equation}
 \left(\IP_{BRST} \,Q_{SUSY} \right) \, \vert \, \mbox{Instanton} \, >
 \,=  \, 0
 \label{inststato}
\end{equation}
In the case of topological field theories the projected supersymmetry
charge is interpreted as the $BRST$--charge $Q_{BRST}$ associated with the
topological symmetry and the generalized instanton configurations
satisfying eq.\eqn{inststato}, being in the kernel of $Q_{BRST}$, are
representatives of the cohomology classes of {\it physical states}.
For the same reason as above \eqn{inststato} are {\it first order
differential equations}. \par
In this lecture I will illustrate the relation between supersymmetry
and the  first order differential equations for either BPS states or
generalized instantons with examples taken from both rigid and local
$N=2$ theories in $D=4$. Here the recently obtained fully general
form of $N=2$ SUGRA and $N=2$ SYM \cite{jgpnoi} allows to match the
structure of Special K\"ahler geometry
\cite{specspec2,skgsugra_4,skgsugra_1} with the structure of the
first order differential equations. In particular a vast number of
results were recently obtained for the case of $N=2$ extremal
black--holes \cite{cardoso,kalvanp1,ferkal2}, \cite{strom3,ferkal4,kalmany},
\cite{carbon}.
The main idea will be reviewed in section ~\ref{BPSlocala}. In sections
~\ref{BPSrigida} and ~\ref{hyperinstanton} the distinction between
the equations for $N=2$ BPS--saturated monopoles   and
those for the gauged hyperinstanton
\cite{topftwist_1,topftwist_2,topf4d_8,witmono} will be
drawn. Finally a short report of work in progress on the application
of solvable Lie algebras to the first order equations for
$BPS$--saturated states in maximally extended supergravities will be
given.
\section{CENTRAL CHARGES}
\label{centcharge}
Let us consider the $D=4$ supersymmetry algebra with an even number
$N=2\nu$ of supersymmetry charges. It can be written in the following
form:
\begin{eqnarray}
&\left\{ {\bar Q}_{Ai \vert \alpha }\, , \,{\bar Q}_{Bj \vert \beta}
\right\}\, = \, & \nonumber\\
&{\rm i} \left( C \, \gamma^a \right)_{\alpha \beta} \,
P_a \, \delta_{AB} \, \delta_{ij} \, - \, C_{\alpha \beta} \,
\epsilon_{AB} \, \times \, \ZZ_{ij}& \nonumber\\
&\left( A,B = 1,2 \qquad ; \qquad i,j=1,\dots, \nu \right)&
\label{susyeven}
\end{eqnarray}
where the SUSY charges ${\bar Q}_{Ai}\equiv Q_{Ai}^\dagger \gamma_0=
Q^T_{Ai} \, C$ are Majorana spinors, $C$ is the charge conjugation
matrix, $P_a$ is the 4--momentum operator, $\epsilon_{AB}$ is the
two--dimensional Levi Civita symbol and the symmetric tensor
$\ZZ_{ij}=\ZZ_{ji}$ is the central charge operator. It
can always be diagonalized $\ZZ_{ij}=\delta_{ij} \, Z_j$ and its $\nu$ eigenvalues
$Z_j$ are the central charges.
\par
The Bogomolny bound on the mass of a generalized monopole state:
\begin{equation}
M \, \ge \, \vert \, Z_i \vert \qquad \forall Z_i \, , \,
i=1,\dots,\nu
\label{bogobound}
\end{equation}
is an elementary consequence of the supersymmetry algebra and of the
identification between {\it central charges} and {\it topological
charges}. To see this it is convenient to introduce the following
reduced supercharges:
\begin{equation}
{\bar S}^{\pm}_{Ai \vert \alpha }=\frac{1}{2} \,
\left( {\bar Q}_{Ai}\, \gamma_0 \, \pm \mbox{i} \, \epsilon_{AB} \,  {\bar Q}_{Bi}\,
\right)_\alpha
\label{redchar}
\end{equation}
They can be regarded as the result of applying
a projection operator to the supersymmetry
charges:
\begin{eqnarray}
{\bar S}^{\pm}_{Ai} &=& {\bar Q}_{Bi} \, \IP^\pm_{BA} \nonumber\\
 \IP^\pm_{BA}&=&\frac{1}{2}\, \left({\bf 1}\delta_{BA} \pm \mbox{i} \epsilon_{BA}
 \gamma_0 \right)
 \label{projop}
\end{eqnarray}
Combining eq.\eqn{susyeven} with the definition \eqn{redchar} and
choosing the rest-frame where $P_a =(M,0,0,0)$ we
obtain the algebra:
\begin{equation}
\left\{ {\bar S}^{\pm}_{Ai}  \, , \, {\bar S}^{\pm}_{Bj} \right\} =
\pm \epsilon_{AC}\, C \, \IP^\pm_{CB} \, \left( M \mp Z_i \right)\,
\delta_{ij}
\label{salgeb}
\end{equation}
By positivity of the operator $\left\{ {S}^{\pm}_{Ai}  \, , \, {\bar S}^{\pm}_{Bj} \right\} $
it follows that on a generic state the Bogomolny bound \eqn{bogobound} is
fulfilled. Furthermore it also follows that the states which saturate
the bounds:
\begin{equation}
\left( M\pm Z_i \right) \, \vert \mbox{BPS state,} i\rangle = 0
\label{bpstate1}
\end{equation}
are those which are annihilated by the corresponding reduced supercharges:
\begin{equation}
{\bar S}^{\pm}_{Ai}   \, \vert \mbox{BPS state,} i\rangle = 0
\label{susinvbps}
\end{equation}
\section{BPS STATES IN RIGID N=2 SUPERSYMMETRY}
\label{BPSrigida}
The most general form of a rigid N=2 super Yang--Mills Lagrangian was derived
in \cite{jgpnoi}: its structure is fully determined by three geometrical data:
\begin{itemize}
\item {The choice of a Special K\"ahler manifold of the rigid type
$ {\cal SK}^{rig}$ describing the vector multiplet couplings}
\item {The choice of a HyperK\"ahler manifold $ {\cal HK}$ describing
the hypermultiplet dynamics }
\item{The choice of a gauge group
$G^{gauge}\subset G^{iso}$,
subgroup of the isometry group of both $ {\cal SK}^{rig}$ and  $ {\cal HK}$ }
\end{itemize}
The bosonic action has the following form:
\begin{eqnarray}
&{\cal L}_{N=2\, SUSY}^{Bose}  =   & \nonumber\\
& g_{i {{j}^\star}}\,
\nabla^{\mu} z^i \nabla _{\mu} \bar z^{{j}^\star}\,+\,
h_{uv} \, \nabla _{\mu}\, q^u \, \nabla^{\mu}\, q^v & \nonumber \\
& - \, {\rm V}\bigl ( z, {\bar
z}, q \bigr ) &
\nonumber\\
&  + \,{\rm i} \,\left(
\bar {\cal N}_{IJ} {\cal F}^{- I}_{\mu \nu}
{\cal F}^{- J\vert {\mu \nu}}
\, - \,
{\cal N}_{IJ} {\cal F}^{+ I}_{\mu \nu}
{\cal F}^{+ J \vert {\mu \nu}} \right )  &
\label{susaction}
\end{eqnarray}
where the scalar potential is expressed in terms of the Killing
vectors $k^i_{I}$, $k^u_{I}$ generating the gauge group algebra on the scalar manifold
${\cal SK}\otimes {\cal HQ}$, of the upper
half $Y^{J}(z)$ of the symplectic section of rigid special geometry and also in terms
of the {\it momentum map} functions ${\cal P}^x_I(q)$ yielding the Poissonian realization
of the gauge group algebra on the HyperK\"ahler manifold:
\begin{eqnarray}
&  \,-{\rm V}\bigl ( z, {\bar
z}, q \bigr )= & \nonumber \\
  & -\,g^2\,\Bigl [ \left( g_{i{{j}^\star}} \, k^i_{I}\,k^{{j}^\star}_{J} +\,4\,h_{uv}
k^u_{I}\,k^v_{J} \right)\,
   \bar Y^{I}\,Y^{J}& \nonumber \\
&  +\,g^{i{{j}^\star}}\,f^{I}_i\,f^{J}_{{j}^\star}\,
{\cal P}^x_{I}\,{\cal P}^x_{J}\,
 \Bigr] &
\label{susypot}
\end{eqnarray}
The kinetic term of the vectors in \eqn{susaction} involves the
period matrix  ${\cal N}_{IJ}$, which is also a datum
of rigid special geometry.
\par
If we restrict our attention to a pure gauge theory without
hypermultiplets, and we calculate the energy of a generic {\it static configuration}
({\it i.e} $F^I_{0a}$ = $0$,  $\nabla_0 z^i$ = $0$), we obtain:
\begin{eqnarray}
& E= \int \, d^3x \Bigl [  \mbox{Im}{\cal N}_{IJ} F^I_{ab}\,  F^J_{ab}
& \nonumber\\
& + \, g_{i{{j}^\star}} \, \nabla_a z^i \,
\nabla_a {\bar z}^{j^\star} \, + \,g^2\, g_{i{{j}^\star}} \, k^i_{I}\,k^{{j}^\star}_{J}\,
   \bar Y^{I}\,Y^{J} \Bigr ]&
\label{monenergy}
\end{eqnarray}
Using the special geometry identity:
\begin{equation}
U^{IJ} \equiv g^{i{{j}^\star}} \, {\bar f}^J_{j^\star} \, f^I_i \, =
\,-\, \frac{1}{2} \, \left ( \mbox{Im} {\cal N} \right)^{-1\vert IJ}
\label{rigident}
\end{equation}
the energy integral \eqn{monenergy} can be rewritten according to a
{\it Bogomolny decomposition} as follows:
\begin{eqnarray}
&E= \int \, d^3x  \, \frac{1}{2}\, g_{i{{j}^\star}}
\left( 2\mbox{i}\, G^i_{ab} \pm \epsilon_{abc} \nabla_c z^i \right)
 & \nonumber\\
&  \times\left(- 2\mbox{i}\, G^{j^\star}_{ab} \pm \epsilon_{abc} \nabla_c z^{j^\star} \right)
&
\label{1stcondi} \\
& + \int \, d^3x  \, \,g^2\, g_{i{{j}^\star}} \, k^i_{I}\,k^{{j}^\star}_{J}\,
   \bar Y^{I}\,Y^{J} & \label{2ndcondi}  \\
 &\pm \int \, d^3x  \, \mbox{i} \epsilon_{abc} \left(G^{j^\star}_{ab} \nabla_c z^{i}
 \, - \,  G^{i}_{ab} \nabla_c z^{j^\star} \right)&
 \label{topocharge}
\end{eqnarray}
where, by definition:
\begin{equation}
G^{i}_{\mu\nu} \, \equiv \, g^{i{{j}^\star}}
\,{\bar f}^I_{j^\star} \,\mbox{Im}{\cal N}_{IJ} F^J_{\mu\nu}   \quad
; \quad
{\bar f}^I_{j^\star} \equiv \nabla_{j^\star}{\bar Y}^I
\end{equation}
The last \eqn{topocharge} of the three addends contributing to the energy is
the integral of a total divergence and can be identified with the
topological charge of the configuration:
\begin{eqnarray}
& Z \, \equiv \, 2 \int_{S_\infty^2} \,\mbox{Im}{\cal N}_{IJ} F^I \, \mbox{Im} Y^J & \nonumber\\
& = 2 \int_{R^3} \, \mbox{Im}{\cal N}_{IJ} F^I \, \wedge \nabla \mbox{Im} Y^J & \nonumber \\
& = \int \, d^3x  \, \mbox{i} \epsilon_{abc} \left(G^{j^\star}_{ab} \nabla_c z^{i}
 \, - \,  G^{i}_{ab} \nabla_c z^{j^\star} \right)\, g_{ij^\star}&
 \label{centratopo}
\end{eqnarray}
where $S_\infty^2$ is the $2$--sphere at infinity bounding  a constant time
slice of space--time. Since the other two addends \eqn{1stcondi},\eqn{2ndcondi} to the energy
of the static configuration are integrals of perfect squares, it follows that in each
topological sector, namely at fixed value of the topological charge
$Z$ the mass satisfies the Bogomolny bound \eqn{bogobound}.
Furthermore a BPS saturated  state (monopole or dyon) is defined by
the two conditions:
\begin{eqnarray}
 &2\mbox{i}\, G^i_{ab} \pm \epsilon_{abc} \nabla_c z^i = 0& \label{BPScondi1}\\
 & g^2\, g_{i{{j}^\star}} \, k^i_{I}\,k^{{j}^\star}_{J}\,
   \bar Y^{I}\,Y^{J} = 0 &
\label{BPScondi2}
\end{eqnarray}
The relation with the preservation of $\frac{1}{2}$ supersymmetries can now be easily seen.
In a bosonic background the supersymmetry variation of the bosons is
automatically zero since it is proportional to the fermion fields
which are zero: one has just to check the supersymmetry variation of
the fermion fields. In the theory under consideration the only
fermionic field is the gaugino and its SUSY variation is given by
(see \cite{jgpnoi}):
\begin{eqnarray}
& \delta\lambda^{iA} \, = \,& \nonumber\\
&\mbox{i} \nabla_\mu z^i \, \gamma^\mu \, \epsilon^A
  + \varepsilon^{AB} \left( G^{-i}_{\mu\nu} \, \gamma^{\mu\nu} + k^i_I {\bar Y}^I \right) \,
\epsilon_B&
\label{gaugvari}
\end{eqnarray}
If we use a SUSY parameter subject to the condition:
\begin{equation}
\gamma_0 \, \epsilon^A \, = \, \pm \mbox{i} \, \varepsilon ^{AB} \,
\epsilon_B
\label{parcondicio}
\end{equation}
then, in a static bosonic background eq.\eqn{gaugvari} becomes:
\begin{eqnarray}
& \delta\lambda^{iA} \, = \,& \nonumber\\
& \Bigl [ -\mbox{i} \,\frac{1}{2} \, \left( 2 \mbox{i}\, G^i_{ab} \pm
\epsilon_{abc} \, \nabla_c z^i \right) \, \gamma^{ab} \, \epsilon_B \,
 & \nonumber \\
& + k^i_I {\bar Y}^I \, \epsilon_B \, \Bigr ]\, \varepsilon^{AB} \,
&
\label{varispec}
\end{eqnarray}
Henceforth the configuration is invariant under the supersymmetries
of type \eqn{parcondicio} if and only if eq.\eqn{BPScondi1} is
satisfied together with:
\begin{equation}
 k^i_I {\bar Y}^I \, = \, 0
 \label{sqrtcondi}
\end{equation}
Eq.\eqn{sqrtcondi} is nothing else but the square--root of eq.\eqn{BPScondi2}.
So we can conclude that the BPS saturated states are just those
configurations which are invariant under supersymmetries of type
\eqn{parcondicio}. On the other hand, these supersymmetries are,
by definition, those generated by
the operators \eqn{redchar}. So by essential use of the {\it
rigid special geometry} structure we have shown the match between
the abstract reasoning of section ~\ref{centcharge} and the
concrete field theory realization of BPS saturated states.
\section{HYPERINSTANTONS IN RIGID $N=2$ SUPERSYMMETRY}
\label{hyperinstanton}
As an exemplification of the alternative case, namely of the relation
between supersymmetry in the Euclidean region with the first order
equations determining generalized {\it instanton configuration}
let us reconsider the complete bosonic lagrangian for a rigid
$N=2$ theory \eqn{susaction} and let us put the vector multiplet
scalars to zero $Y^I=0$, while keeping the hypermultiplet ones.
This choice yields the gauge--fixed, ghost--free sector of the
topological field theory associated with the considered $N=2$ theory.
Indeed, when applying the topological twist procedure, as
defined in \cite{topftwist_1,topftwist_2}, the physical fields
are identified with the gauge bosons $A_\mu ^I$ and with the hypermultiplet
scalars $q^u$, while all the other fields, including all the fermions
and the vector multiplet scalars are ghosts. Explicitly the
ghost--free lagrangian, rotated to Euclidean space, is:
\begin{eqnarray}
{\cal L}^E & =&  -\mbox{i}\Bigl( {\bar {\cal N}}_{IJ} \,
F^{-I}_{\mu\nu} \,F^{-J}_{\mu\nu}
 -{{\cal N}}_{IJ}\,F^{+I}_{\mu\nu} \,F^{+J}_{\mu\nu}
\Bigr) \nonumber\\
&& +h_{uv} \,\nabla_\mu q^u \,\nabla_\nu q^v  \nonumber\\
&& + g^{ij^\star} \,
f^I_{i} \, {\bar f}^J_{j^\star} \, \sum_{x=1}^{3} \, {\cal P}^x_I \, {\cal P}^x_J
\label{euclid}
\end{eqnarray}
Using the identities of rigid special geometry (see \cite{jgpnoi}),
the Euclidean action $S^E=\int \, L^E \, d^4x$ can be rewritten as:
\begin{eqnarray}
S^E &=& \mbox{i}\, \frac{1}{2}\, {\cal N}_{IJ} \, \int \, F^I \wedge
F^J  \label{topol1}\\
&&- \, \frac{1}{2} \, \sum_{x=1}^{3} \int {\hat K}^x \wedge \Theta^{x-}
\label{topol2} \\
&&+ \int \, d^4x \left[ 2 \vert \vert B^-\vert \vert^2 +\frac{1}{4}
\vert \vert b^-\vert \vert^2 \right]
\label{hydecomp}
\end{eqnarray}
where the first two terms are topological integrals, the first being
a Chern class, the second the product of the {\it gauged
HyperK\"ahler class} of the target manifold (pull--back thereof):
\begin{equation}
{\hat K}^x = \left(K^x_{uv} \, \nabla_\mu q^u \nabla_\nu q^v \, +\, {\cal P}^x_I \,
{\cal F}^I_{\mu \nu}\right)\, dx^\mu  \wedge dx^\nu
\label{HyKclass}
\end{equation}
with the  triplet of antiself--dual 2--forms $\Theta^{x-}$, satisfying
the quaternionic algebra, that are defined on the world manifold, namely
on Euclidean space--time. From the topological field theory viewpoint
the first two addends to the Euclidean action \eqn{topol1} and \eqn{topol2}
constitute the classical action. The remaining two terms \eqn{hydecomp}
that are perfect squares constitute the gauge fixing terms. The gauge
fixing conditions are the following two {\it first order differential
equations}:
\begin{eqnarray}
0 &=& B^{I-}_{\mu \nu} \equiv F^{-I}_{\mu\nu} +\nonumber\\
&&\left(\mbox{Im}{\cal
N}\right)^{-1\vert IJ}\, \sum_{x=1}^{3} \,\Theta^{x-}_{\mu \nu} \, {\cal P}^x_J
\label{hyper1}\\
0 &=& b^{-u}_\mu \equiv \nabla_\mu q^u - \nonumber\\
&& \sum_{x=1}^{3} \,
\Theta^{x-}_{\mu \nu} \, \nabla_\rho q^v \, K^x_{vw} \, h^{wv} \,
g^{\nu\rho}\label{hyper2}
\end{eqnarray}
Eq.s \eqn{hyper1} and \eqn{hyper2} correspond to a generalization of
the classical instanton equations of Yang--Mills theory. Rather than
stating that the field strength is self--dual eq.\eqn{hyper1} relates its
antiself dual part to {\it an algebraic expression in the matter
fields} (=the momentum map ${\cal P}^x_J$). At the same time the map
\begin{equation}
q: \, {\cal M}_{world} \, \longrightarrow \, {\cal M}_{target}
\end{equation}
from space--time to the target manifold is constrained to be a {\it
triholomorphic map} by eq.\eqn{hyper2}. Indeed this {\it first order
equation} is a generalization to curved manifolds of the {\it Cauchy
Fueter} equations, that is the quaternionic version of {\it Cauchy
Riemann equations} for holomorphic maps.
\par
Field configurations satisfying the first order equations \eqn{hyper1},
\eqn{hyper2} were named {\it gauged hyperinstanton} by the present
author in a series of papers written in collaboration with D. Anselmi
\cite{topftwist_1,topftwist_2,topf4d_8}. At fixed topological
numbers (=the value of the integrals \eqn{topol1},\eqn{topol2} ) the
hyperinstanton configuration corresponds to the absolute minimum of
the Euclidean action. Furthermore in
the topological field theory  the functional integral
is reduced to an integral over the
{\it moduli space of hyperinstantons}. Indeed in the topological field
theory each hyperinstanton
is  a representative of an entire orbit of the classical symmetry group,
that is the deformation group of connections and maps. Hence reducing the path
integral to orbit space we obtain a summation over hyperinstantons.
\par
As shown in \cite{topftwist_1,topftwist_2,topf4d_8} and further
discussed in \cite{noi}, eq.s \eqn{hyper1},\eqn{hyper2} are obtained
by setting to zero the SUSY variation of the {\it gaugino} and of the {\it
hyperino} with respect to a  certain combination of  the $8$ SUSY parameters
$\epsilon^{\alpha A}$.  This combination is derived in the following
way. In the Euclidean region the Lorentz group $SO(1,3)$ becomes
$SU(2)_L \otimes SU(2)_R$ which makes a triplet os $SU(2)$ groups with
$SU(2)_I$, the automorphism group of the $N=2$ SUSY algebra. With
respect to the group $SU(2)_L \otimes SU(2)_R \otimes SU(2)_I$ the
parameters $\epsilon^{\alpha A}$ fall in the following
representations
\begin{equation}
\epsilon^{\alpha A} =  \left(\frac{1}{2},0,\frac{1}{2}\right)\, \oplus \,
\left(0,\frac{1}{2},\frac{1}{2}\right)\,
\end{equation}
The appropriate combination of SUSY parameters used in this game
is the component of $\epsilon^{\alpha A}$ which is
a singlet with respect to the diagonal $SU(2)_{diag}$ =
$\mbox{diag} \left(SU(2)_L\otimes SU(2)_I \right)$.
The topological field theory interpretation of this fact is that,
after topological twist, the corresponding SUSY generator is
reinterpreted as BRTS charge and the gaugino and hyperino fields
of a certain chirality as antighosts. Hence their BRST variation is,
by definition, the gauge--fixing condition.
\par
In the current literature, the hyperinstanton eq.s
\eqn{hyper1},\eqn{hyper2} are often named
Seiberg--Witten monopole equations. Yet it should be noted
that they are instanton--like and not monopole equations.
Indeed
they occur in the Euclidean version of $4$--dimensional space--time and
not in $3$--space. They are obtained from the afore mentioned combination
of supersymmetries
and not from the projected supersymmetry of eq.\eqn{parcondicio}.
They were originally discovered by the present author with Anselmi
\cite{topftwist_1,topftwist_2}
in the more general local supersymmetry context. They were applied by
Witten in his paper on Monopoles and 4--manifolds where they
were utilized to calculate Donaldson invariants advocating the
results of Seiberg--Witten theory \cite{seiwit}.
\section{BPS BLACK HOLES IN  N=2 LOCAL SUPERSYMMETRY}
\label{BPSlocala}
Eq.\eqn{parcondicio} is not Lorentz invariant and introduces a
clear--cut separation between space and time. The interpretation of
this fact is that we are dealing with localized lumps of energy that can be interpreted as
quasi--particles at rest. A Lorentz boost simply puts such quasi--particles
into motion. In the gravitational case the generalization of eq.\eqn{parcondicio}
requires the existence of a time--like killing vector $ \xi^\mu$, in order
to write:
\begin{equation}
 \xi^\mu \,\gamma_\mu \, \epsilon^A \, = \, \pm \mbox{i} \, \varepsilon ^{AB} \,
\epsilon_B
\label{lparcondicio}
\end{equation}
Furthermore, the analogue of the localization condition corresponds
to the {\it asymptotic flatness} of space--time.
\par
We are therefore led to look for the BPS saturated states of local $N=2$
supersymmetry within the class of {\it electrically and magnetically charged,
asymptotically flat, static space--times}. Generically such
space--times are black--holes since they have singularities hidden by horizons.
  Without the constraints imposed by supersymmetry the
horizons can also disappear and there exist configurations that display
{\it naked singularities}. In the supersymmetric case, however, the
Bogomolny bound \eqn{bogobound} becomes the statement that the ADM
mass of the black--hole is always larger or equal than the central charge.
This condition just ensures that the horizon exists. Hence the
{\it cosmic censorship} conjecture is just a consequence of $N\ge 2$
supersymmetry. This was noted for the first time in \cite{kalvanp1}.
The BPS saturated black--holes are configurations for
which the horizon area is minimal at fixed electric and magnetic
charges. This result was obtained by Ferrara and Kallosh in
\cite{ferkal4,ferkal2}.They are determined by solving the gravitational analogue
of the Bogomolny first order equations \eqn{BPScondi1}, \eqn{sqrtcondi},
obtained from the SUSY variation of the fermions.
\par
If we restrict our attention to the gravitational coupling of vector
multiplets, the bosonic action we have to consider is the
following one:
 \begin{eqnarray}
&S^{Bose}_{N=2} =
 \int \sqrt{-g}\,d^4 \,x  {\cal L}& \nonumber \\
& {\cal L}    =
  - {1 \over 2} \, R \,  + \, g_{i {{j}^\star}}\,
\nabla^{\mu} z^i \nabla _{\mu} \bar z^{{j}^\star}   &
\nonumber\\
&  + \,{\rm i} \,\left(
\bar {\cal N}_{\Lambda \Sigma} {\cal F}^{- \Lambda}_{\mu \nu}
{\cal F}^{- \Sigma \vert {\mu \nu}}
\, - \,
{\cal N}_{\Lambda \Sigma} {\cal F}^{+ \Lambda}_{\mu \nu}
{\cal F}^{+ \Sigma \vert {\mu \nu}} \right )  &
 \nonumber    \\
&  -\,g^2\, g_{i{{j}^\star}} \, k^i_{\Lambda}\,k^{{j}^\star}_{\Sigma}
\, \bar L^{\Lambda}\,L^{\Sigma} &
\label{sugraction}
\end{eqnarray}
where $L^{\Lambda}$ denotes the upper half of a covariantly
holomorphic section of {\it local special geometry} and
${\cal N}_{\Lambda \Sigma}$ is the period matrix according to
its local rather than rigid definition (see \cite{mylec},\cite{jgpnoi}).
According to the previous discussion we consider for the metric an ansatz
of the following form:
\begin{equation}
ds^2 \, = \, e^{2U(r)} \, dt^2 \, - \, e^{-2U(r)} d{\vec x}^2
\label{umetric}
\end{equation}
where ${\vec x}$ are isotropic coordinates on $\IR^3$ and
$U(r)$ is a function only of:
\begin{equation}
r \equiv \sqrt{{\vec x}^2}
\label{rfun}
\end{equation}
As we shall see in the next section, eq.\eqn{umetric} corresponds to
a 0--brane ansatz. This is in line with the fact that we have 1--form
gauge fields in our theory that couple to 0--branes, namely to particle
world--lines. Indeed, in order to proceed further we need an ansatz for
the gauge field strengths. To this effect we begin by constructing a
2--form which is {\it anti--self--dual} in the background of the
metric \eqn{umetric} and whose integral on the $2$--sphere at
infinity $ S^2_\infty$ is normalized to $ 2 \pi $. A short
calculation yields:
\begin{eqnarray}
E^-   &=&   \mbox{i} \frac{e^{2U(r)}}{r^3} \, dt \wedge {\vec x}\cdot
d{\vec x}  +  \frac{1}{2}   \frac{x^a}{r^3} \, dx^b   \wedge
dx^c   \epsilon_{abc} \nonumber\\
2 \, \pi  &=&    \int_{S^2_\infty} \, E^-
\label{eaself}
\end{eqnarray}
and with a little additional effort one obtains:
\begin{equation}
E^-_{\mu\nu} \, \gamma^{\mu\nu}  = 2 \,\mbox{i} \frac{e^{2U(r)}}{r^3}  \,
\gamma_a x^a \, \gamma_0 \, \frac{1}{2}\left[ {\bf 1}+\gamma_5 \right]
\label{econtr}
\end{equation}
which will prove of great help in the unfolding of the supersymmetry
transformation rules. Then utilizing eq.\eqn{eaself} we write the
following ansatz for the gauge field--strengths:
\begin{eqnarray}
F^{\Lambda -}_{\mu\nu} & \equiv & \frac{1}{2} \left(F^{\Lambda }_{\mu\nu} -
\mbox{i}\,\frac{1}{2}\, \epsilon_{\mu\nu\rho\sigma} \,F^{\Lambda \vert\rho\sigma }
\right)
\nonumber \\
 & =& \frac{1}{4 \pi}\, t^\Lambda \, E^-_{\mu\nu} \nonumber\\
 t^\Lambda & = & \mbox{complex number} \nonumber\\
            & =&\int_{S^2_\infty}\,  F^{\Lambda -}_{\mu\nu} \, dx^\mu
            \wedge \, dx^\nu
 \label{gaugansz}
\end{eqnarray}
 Following the standard definitions occurring in the discussion of electric--magnetic
 duality rotations \cite{mylec} (and \cite{jgpnoi}) we also obtain:
\begin{eqnarray}
G^{-}_{\Lambda\vert \mu\nu}& = &{\bar {\cal N}}_{\Lambda\Sigma}
\, F^{\Sigma -}_{\mu\nu}\nonumber\\
& = & \frac{1}{4 \pi} \,{\bar {\cal N}}_{\Lambda\Sigma}\, t^\Sigma \, E^-_{\mu\nu}
\label{Gform}
\end{eqnarray}
To our purposes the most important field strength combinations are
the gravi--photon and matter--photon combinations occurring, respectively
in the gravitino and gaugino SUSY rules. They are defined by
(see \cite{jgpnoi}):
\begin{eqnarray}
 T^{-}_{\mu\nu}& =& 2\mbox{i}\, \left( \mbox{Im}
 {\cal N}\right)_{\Lambda\Sigma} \, L^\Lambda \,F^{\Sigma -}_{\mu\nu}
 \label{gravifot} \\
 G^{\pm i}_{\mu\nu} & = & -g^{ij^\star} \, {\bar f}^\Gamma_{j^\star}
 \, \left( \mbox{Im} {\cal N}\right)_{\Gamma\Lambda} \,F^{\Sigma \pm}_{\mu\nu}
 \label{matfot}
\end{eqnarray}
The central charge and the matter charhes are defined by the integral of the graviphoton
\eqn{gravifot} (see \cite{cereferpro}):
\begin{eqnarray}
Z & \equiv & \int_{S^2_\infty} \,T^{-}_{\mu\nu} \, dx^\mu \wedge dx^\nu \nonumber\\
Z^i & \equiv & \int\, G^{\pm i}_{\mu\nu} \, dx^\mu \wedge dx^\nu \nonumber\\
&& = g^{ij^\star} \, \nabla_{j^\star} {\bar Z}
\label{zdef}
\end{eqnarray}
Using eq.\eqn{gaugansz} and \eqn{gravifot}we obtain:
\begin{equation}
Z= 2\mbox{i}\, \left( \mbox{Im}
 {\cal N}\right)_{\Lambda\Sigma} \, L^\Lambda \, t^\Sigma
 \label{espres}
\end{equation}
while utilizing the identities of special geometry we also obtain:
\begin{equation}
T^{-}_{\mu\nu} = M_\Sigma \, F^\Sigma_{\mu\nu} - L^\Lambda \,
G_{\Lambda \vert \mu\nu}
\label{passaggio}
\end{equation}
where $M_\Sigma(z)$ is the lower part of the symplectic section of local
special geometry. Consequently we obtain:
\begin{eqnarray}
Z= M_\Sigma \, p^\Sigma - L^\Lambda \, q_\Lambda
\label{holcharge}
\end{eqnarray}
having defined the moduli dependent electric and magnetic
charges as follows:
\begin{eqnarray}
q_\Lambda & \equiv & \int_{S^2_\infty} \,G_{\Lambda\vert \mu\nu}\,
dx^\mu \wedge dx^\nu
\label{holele}\\
p^\Sigma & \equiv & \int_{S^2_\infty} \,F^{\Sigma}_{ \mu\nu}\,\,
dx^\mu \wedge dx^\nu
\label{holmag}
\end{eqnarray}
Alternatively, if following J. Schwarz \cite{sumschwarz} we define the
electric and magnetic charges by the asymptotic behaviour  of the
bare electric and magnetic fields:
\begin{equation}
F^{\Lambda}_{0a} \cong \frac{q^\Lambda_{(el)}}{r^3}\, x^a \quad ; \quad
{\tilde F}^{\Lambda}_{0a} \cong \frac{q^\Lambda_{(mag)}}{r^3}\, x^a
\label{elemag}
\end{equation}
we find the relations
\begin{equation}
q^\Lambda_{(el)}= 2\, \mbox{Im}\, t^\Lambda \quad ; \quad  q^\Lambda_{(mag)}
= 2\, \mbox{Re}\, t^\Lambda
\label{relazia}
\end{equation}
and
\begin{eqnarray}
t^\Lambda&=&\frac{1}{2}\,\Biggl \{ p^\Lambda+\mbox{i}\left(\mbox{Im} {\cal
N}_\infty^{-1}\right)^{\Lambda\Sigma} \nonumber\\
&& \times \, \left[\left(\mbox{Re}{\cal
N}\right)_{\Sigma\Gamma}\, p^\Gamma - q_\Sigma\right]\Biggr \}
\label{relaziona}
\end{eqnarray}
In a fully general bosonic background the $N=2$ supersymmetry
transformation rules of the gravitino and of the gaugino are:
\begin{eqnarray}
\delta\,\psi_{A \vert \mu} &=& \nabla_\mu \epsilon_A -
\frac{1}{4} T^-_{\rho\sigma} \gamma^{\rho\sigma}
\, \gamma _\mu \,\epsilon^B \, \varepsilon ^{AB} \label{gravirule}\\
\delta\lambda^{\alpha A} & = & \mbox{i} \, \nabla_\mu z^\alpha \,
\gamma^\mu \, \epsilon^A + G^{-\alpha}_{\rho\sigma}
\gamma^{\rho\sigma} \epsilon^B \, \varepsilon ^{AB}
\nonumber\\
&&+ \varepsilon ^{AB} \, k^\alpha_\Lambda \, {\bar L}^\Lambda \,
\epsilon_B \label{gaugirule}
\end{eqnarray}
where the derivative:
\begin{equation}
\nabla_\mu \epsilon_A   \equiv   \Bigl (\partial_\mu -\frac{1}{4}
\omega^{ab}_\mu \, \gamma_{ab}
 +\mbox{i}\,\frac{1}{2} Q_\mu \Bigr ) \epsilon_A
\label{covloka}
\end{equation}
is covariant both with respect to the Lorentz and with
respect to the K\"ahler transformations. Indeed it also contains
the K\"ahler connection:
\begin{equation}
Q_\mu  \equiv - \mbox{i}\, \frac{1}{2} \left(\partial_i {\cal
K}\partial_\mu z^i - \partial_{i^\star} {\cal
K}\partial_\mu {\bar z}^{i^\star} \right)
\label{kacon}
\end{equation}
As supersymmetry parameter we choose one of the following form:
\begin{eqnarray}
\epsilon_A &=& e^{f(r)} \chi _A \quad \mbox{$ \chi_A $ = constant and}\nonumber\\
           && \gamma_0 \chi ^A = \pm \,\mbox{i} \, \varepsilon^{AB} \, \chi _B
           \label{susyparam}
\end{eqnarray}
Using the explicit form of the spin connection for the metric
\eqn{umetric}:
\begin{eqnarray}
\omega^{0a} &=& -\, \partial_a U \, dt \, e^{2U}\nonumber\\
\omega^{ab} &=& 2 \, \partial^a U \, dx^b
\label{spincon}
\end{eqnarray}
and inserting the SUSY parameter \eqn{susyparam} into the
gravitino variation \eqn{gravirule}, from the invariance
condition $ \delta \psi _{A \vert \mu} = 0$ we obtain two equations
corresponding respectively to the case $\mu = 0 $ and to the case
$\mu = a $. Explicitly we get:
\begin{eqnarray}
\frac{dU}{dr} & = & \mp \, 2 \,\mbox{i}\,
\left(\mbox{Im}{\cal  N}\right)_{\Lambda\Sigma}
\, L^\Lambda t^\Sigma \, \frac{e^U}{r^2}
\label{Uequa}\\
\frac{df}{dr} &= &-  \frac{1}{2} \, \frac{dU}{dr}  +  \mbox{i}\frac{1}{2} \,  \Bigl (
\partial_i {\cal K}\frac{d z^i}{dr} - \nonumber\\
&&\partial_{i^\star} {\cal K}\frac{d{\bar z}^{i^\star}}{dr}
\Bigr )\label{fequa}
\end{eqnarray}
 On the other hand setting to zero the gaugino transformation rule
 \eqn{gaugirule} with the SUSY parameter \eqn{susyparam} we obtain:
\begin{equation}
\frac{dz^i}{dr}=\mp 2 \, \mbox{i}\, g^{ij^\star}{\bar f}^\Lambda_{j^\star}
\left( \mbox{Im}{\cal N}\right)_{\Lambda\Sigma} \, \frac{t^\Sigma}{r^2} \, e^U
\label{zequa}
\end{equation}
In obtaining these results, crucial use was made of eq.\eqn{econtr}.
\par
In this way we have reduced the equations for the extremal
BPS saturated black--holes to a pair of first order differential
equations for the metric scale factor $U(r)$ and for the
scalar fields $z^i(r)$. To obtain explicit solutions one should
specify the special K\"ahler manifold one is working with, namely
the specific Lagrangian model. There are, however, some very
general and interesting conclusions that can be drawn in a
model--independent way. They are just consequences of the fact that
the black--hole equations are {\it first order differential equations}.
Because of that there are fixed points
(see the papers \cite{ferkal2,ferkal4,strom3}) namely values either of the metric
or of the scalar fields which, once attained in the evolution
parameter $r$ (= the radial distance ) persist indefinitely. The
fixed point values are just the zeros of the right hand side in
either of the coupled eq.s \eqn{Uequa} and \eqn{zequa}. The fixed
point for the metric equation is $r=\infty $, which correpsonds to its
asymptotic flatness. The fixed point for the moduli is $r=0$.  So, independently
from the initial data at  $r=\infty$ that determine the details
of the evolution,  the scalar fields flow into
their fixed point values at $r=0$, which, as I will show,
turns out to be a horizon. Indeed in the vicinity of $r=0$ also the
metric takes a universal form.
\par
Let us see this more closely.
\par
To begin with we consider the equations determining the fixed point
values for the moduli and the universal form attained by the metric
at the moduli fixed point:
\begin{eqnarray}
0 &=& -g^{ij^\star} \, {\bar f}^\Gamma_{j^\star}
\left( \mbox{Im}{\cal N}\right)_{\Gamma\Lambda} \, F^{\Lambda -}_{\mu \nu}
\label{zequato} \\
\frac{dU}{dr} & = & \mp \, 2 \,\mbox{i}\,
\left(\mbox{Im}{\cal  N}\right)_{\Lambda\Sigma}
\, L^\Lambda q^\Sigma \, \frac{e^U}{r^2}
\label{Uequato}
\end{eqnarray}
Multiplying eq.\eqn{zequato} by $f^\Sigma_i$, using the local
special geometry counterpart of eq.\eqn{rigident}:
\begin{equation}
 f^\Sigma_i \, g^{ij^\star} \, {\bar f}^\Gamma_{j^\star}= -\frac{1}{2}\,
 \left(\mbox{Im}{\cal N}\right)^{-1\vert \Sigma\Gamma } -
 {\bar L}^\Sigma \, L^\Gamma
 \label{locident}
\end{equation}
and the definition \eqn{gravifot} of the graviphoton field strength
we obtain:
\begin{equation}
0 = -\frac{1}{2} \, F^{\Lambda -}_{\mu \nu} +\mbox{i}\frac{1}{2} \,
{\bar L}^\Lambda \, T^{-}_{\mu \nu}
\label{ideft}
\end{equation}
Hence, using the definition of the central charge
\eqn{zdef} and eq.\eqn{gaugansz} we conclude that
at the fixed point the following condition is true:
\begin{equation}
0=-\frac{1}{2} \, \frac{t^\Lambda}{4\pi} \, -\, \frac{Z_{fix} \,
{\bar L}_{fix}^\Lambda}{8\pi}
\label{passetto}
\end{equation}
In terms of the previously defined electric and magnetic charges
eq.\eqn{passetto} can be rewritten as:
\begin{eqnarray}
p^\Lambda & = & \mbox{i}\left( Z_{fix}\,{\bar L}^\Lambda_{fix}
- {\bar Z}_{fix}\,L^\Lambda_{fix} \right)\\
q_\Sigma & = & \mbox{i}\left( Z_{fix}\,{\bar M}_\Sigma^{fix}
- {\bar Z}_{fix}\,M_\Sigma^{fix} \right)\\
Z_{fix} &=& M_\Sigma^{fix} \, p^\Lambda\, - L^\Lambda_{fix} \, q_\Lambda
\label{minima}
\end{eqnarray}
which can be regarded as algebraic equations determining the value
of the scalar fields at the fixed point as functions of the electric
and magnetic charges $p^\Lambda, q_\Sigma$:
\begin{equation}
L_{fix}^\Lambda = L^\Lambda(p,q) \, \longrightarrow \,
Z_{fix}=Z(p,q)=\mbox{const}
\end{equation}
\par
In the vicinity of the fixed point the differential equation for the
metric becomes:
\begin{equation}
\pm \, \frac{dU}{dr}=\frac{Z(p,q)}{4\pi \, r^2} \, e^{U(r)}
\end{equation}
which has the approximate solution:
\begin{equation}
\exp[U(r)]\, {\stackrel{r \to 0}{\longrightarrow}}\, \mbox{const} + \frac{Z(p,q)}{4\pi \, r}
\label{approxima}
\end{equation}
Hence, near $r=0$   the metric \eqn{umetric}
becomes of the Bertotti Robinson type:
\begin{eqnarray}
ds^2_{BR} &=& \frac{r^2}{m_{BR}^2}\, dt^2 \, - \, \frac{m_{BR}^2}{r^2} \,
dr^2 \, \nonumber\\
&&- \, m_{BR}^2 \, \left(Sin^2\theta \, d\phi^2 + d\theta^2 \right)
\label{bertrob}
\end{eqnarray}
with Bertotti Robinson mass given by:
\begin{equation}
m_{BR}^2 = \vert \frac{Z(p,q)}{4\pi} \vert^2
\label{brmass}
\end{equation}
In the metric \eqn{bertrob} the surface $r=0$ is light--like and
corresponds to a horizon since it is the locus where the
Killing vector generating time translations $\frac{\partial}{\partial t} $,
which is time--like at spatial infinity $r=\infty$, becomes
light--like. The horizon $r=0$ has a finite area given by:
\begin{equation}
\mbox{Area}_H = \int_{r=0} \, \sqrt{g_{\theta\theta}\,g_{\phi\phi}}
\,d\theta \,d\phi \, = \,  4\pi \, m_{BR}^2
\label{horiz}
\end{equation}
Hence, independently from the details of the considered model,
the BPS saturated black--holes in an N=2 theory have a
Bekenstein--Hawking entropy given by the following horizon area:
\begin{equation}
 \mbox{Area}_H = \, \frac{1}{4\pi} \, \vert Z(p,q) \vert^2
 \label{ariafresca}
\end{equation}
the value of the central charge being determined by eq.s
\eqn{minima}. Such equations can also be seen as the variational
equations for the minimization of the horizon area as
given by \eqn{ariafresca}, if the central charge is regarded
as a function of both the scalar fields and the charges:
\begin{eqnarray}
 \mbox{Area}_H (z,{\bar z})&=& \, \frac{1}{4\pi} \, \vert Z(z,{\bar z},p,q) \vert^2
 \nonumber\\
 \frac{\delta \mbox{Area}_H }{\delta z}&=&0 \, \longrightarrow \, \nonumber \\
  z&=&z_{fix}
\end{eqnarray}
\section{WORK IN PROGRESS:
THE $p$--BRANES OF STRING THEORY AND $M$--THEORY AND SOLVABLE LIE
ALGEBRAS}
When $M$--theory (namely the still undefined quantum theory whose
low energy limit is $11$--dimensional supergravity) is compactified
on a torus $T^{r+1}$ or $10$--dimensional string theory is compactified
on a torus $T^{r}$, the low energy interactions of the massless modes are
described by {\it maximal supergravity} in dimensions $D=10-r$. As
discussed in a recent series of papers \cite{solvab1},\cite{solvab2}.
It has been known for many years \cite{sase} that the scalar field manifold of
both pure and matter coupled
$N>2$ extended supergravities in $D=10-r$ ($r=6,5,4,3,2,1$) is a non compact
homogeneous symmetric manifold $U_{(D,N)} /H_{(D,N)}$, where $U_{(D,N)}$ (depending
on the space--time dimensions and on the number of supersymmetries) is a non compact
Lie group and $H_{(D,N)}\subset U_{(D,N)}$ is a maximal compact subgroup.
Furthermore, the structure of the supergravity lagrangian is completely encoded in the
local differential geometry of
$U_{(D,N)}/H_{(D,N)}$, while an appropriate restriction to integers $U_{(D,N)}(\ZZ)$ of
the Lie group $U_{(D,N)}$
is the conjectured U--duality symmetry of string theory that unifies  T--duality with S--duality \cite{huto}.
\par
As we discussed in a recent paper \cite{solvab1}, utilizing a well established mathematical
framework \cite{helgason}, in all these cases the scalar coset manifold $U/H$ can be
identified with the group manifold of a normed solvable Lie algebra:
\begin{equation}
  U/H \sim \exp[{Solv}]
\end{equation}
\par
The representation of the supergravity scalar manifold ${\cal M}_{scalar}= U/H$
as the group manifold associated with a {\it  normed solvable Lie algebra}
introduces a one--to--one correspondence between the scalar fields $\phi^I$ of
supergravity and the generators $T_I$ of the solvable Lie algebra $Solv\, (U/H)$.
Indeed the coset representative $L(U/H)$ of the homogeneous space $U/H$ is
identified with:
\begin{equation}
L(\phi) \, =\, \exp [ \phi^I \, T_I ]
\label{cosrep1}
\end{equation}
where $\{ T_I \}$ is a basis of $Solv\, (U/H)$.
\par
As a consequence of this fact the tangent bundle to the scalar manifold $T{\cal M}_{scalar}$
is identified with the solvable Lie algebra:
\begin{equation}
T{\cal M}_{scalar} \, \sim \,Solv \, (U/H)
\label{cosrep2}
\end{equation}
and any algebraic property of the solvable algebra has a corresponding
physical interpretation in
terms of string theory massless field modes.
\par
Furthermore, the local differential geometry of the scalar manifold is described
 in terms of the solvable Lie algebra structure.
 Given the euclidean scalar product on $Solv$:
\begin{eqnarray}
  <\, , \, > &:& Solv \otimes Solv \rightarrow \IR
\label{solv1}\\
<X,Y> &=& <Y,X>\label{solv2}
\end{eqnarray}
the covariant derivative with respect to the Levi Civita connection is given by
the Nomizu operator \cite{alex}:
\begin{equation}
\forall X \in Solv : \IL_X : Solv \to Solv
\end{equation}
\begin{eqnarray}
   &\forall X,Y,Z \in Solv  :   2 <Z,\IL_X Y>&   \nonumber\\
&=   <Z,[X,Y]> - <X,[Y,Z]>& \nonumber\\
&  -  <Y,[X,Z]> &
\label{nomizu}
\end{eqnarray}
and the Riemann curvature 2--form is given by the commutator of two Nomizu
operators:
\begin{equation}
 <W,\{[\IL_X,\IL_Y]-\IL_{[X,Y]}\}Z> = R^W_{\ Z}(X,Y)
\label{nomizu2}
\end{equation}
In the case of maximally extended supergravities in $D=10-r$ dimensions the scalar
manifold has a universal structure:
\begin{equation}
 { U_D\over H_D}  = {E_{r+1(r+1)} \over H_{r+1}}
\label{maximal1}
\end{equation}
where the Lie algebra of the $U_D$--group $E_{r+1(r+1)} $ is the
maximally non compact real section of the exceptional $E_{r+1}$ series  of the simple complex
Lie Algebras
and $H_{r+1}$ is its maximally compact subalgebra \cite{cre}.
As we discussed in a recent paper \cite{solvab1},
the manifolds $E_{r+1(r+1)}/H_{r+1}$
share the distinctive  property of being non--compact homogeneous spaces of maximal rank
$r+1$, so that the associated solvable Lie algebras,
 such that ${E_{r+1(r+1)}}/{H_{r+1}} \, = \, \exp \left [ Solv_{(r+1)} \right ]
$,  have the particularly simple structure:
\begin{equation}
Solv\, \left ( E_{r+1}/H_{r+1} \right )\, = \, {\cal H}_{r+1} \, \oplus_{\alpha \in
\Phi^+(E_{r+1})} \, \IE^\alpha
\label{maxsolv1}
\end{equation}
where $\IE^\alpha \, \subset \, E_{r+1}$ is the 1--dimensional subalgebra associated
with the root $\alpha$
  $\Phi^+(E_{r+1})$ is the positive part of the $E_{r+1}$--root--system and
 ${\cal H}_{r+1}$ denotes the Cartan subalgebra.
\par
As shown in \cite{solvab2} the solvable Lie algebras $Solv\, \left ( E_{r+1}/H_{r+1} \right )$
can be decomposed in a sequential  way. Indeed we can write the equation:
\begin{equation}
\oplus_{k=1}^{r } \IR \cdot Y_k ={\cal H}_{r} \subset
{\cal H}_{r+1}=\oplus_{k=1}^{r+1} \IR \cdot Y_k
\label{decompo1}
\end{equation}
where $\IR \cdot Y_k$ is the $1$--dimensional vector space associated
with the Cartan generator $Y_k$.
On the other hand  $\Phi^{+}(E_{r+1})$, namely the positive part of the $E_{r+1}$ root space
is split as follows:
\begin{equation}
\Phi^{+}(E_{r+1}) = \Phi^+(E_2) \oplus_{k=1}^{k=r} \ID^{+}_{k+1}
\label{decompo}
\end{equation}
where $ \Phi^+(E_2) $ is the one--dimensional root space of the
U--duality group in $D=9$ and $\ID^{+}_{k+1}$ are the weight-spaces
of the $E_{k+1}$ irreducible
representations to which the vector field in $D=10-r$
are assigned. Alternatively, as it is explained in \cite{solvab2}, ${\cal A}_{r+2}
\equiv \ID^{+}_{r+1}$ are
the maximal abelian ideals of the U--duality group   $E_{r+2}$ in
$D=10-r-1$ dimensions.
\par
Adopting the short hand notation:
\begin{eqnarray}
 \phi \cdot  {\cal H} & \equiv & \phi^i  \,   Y_i  \nonumber\\
 \tau_k \cdot \ID_k & \equiv & \tau^i_{k}  \,D^{(k)}_{i} \nonumber\\
 \label{fucili}
\end{eqnarray}
the coset representative for maximal supergravity in dimension
$D=10-r$ can be written as:
\begin{eqnarray}
L & = & \exp \left [ \phi \cdot  {\cal H} \right ] \,  \prod_{k=1}^{r}
\, \exp \left[   \tau_k \cdot \ID_k \right ] \nonumber\\
& = & \prod_{j=1}^{r+1} S^i \,  \prod_{k=1}^{r}   \,
\left( {\bf 1} + \tau_r \cdot \ID_r \right)
\end{eqnarray}
The last line
follows from the abelian nature of the ideals $\ID_k$ and from the position:
\begin{equation}
S^i  \, \equiv \, \exp[\phi^i Y_i]
\end{equation}
Hence the solvable Lie algebra structure provides a canonical
parametrization of the scalar field manifold where the fields
associated with the Cartan generators are the {\it generalized
dilatons} which appear in the lagrangian in an exponential way, while
the fields associated with the nilpotent generators
appear in the lagrangian only through polynomials of degree bounded
from above.
\par
Since the fermion transformation rules and the associated central charges
of all maximally extended supergravities are expressed solely in
terms of the coset representative $L\left(\phi\right)$ (see \cite{voialtri}),
the  method for the derivation of
extremal  solutions, which in the previous section was applied to the
case of $D=4, N=2$ black--holes, can now be extended to the case of
extremal $p$--brane solutions in $D=10-r$. The canonical parametrization
of the scalars through solvable Lie algebras hints to a complete
solubility of the corresponding first order equations, namely of the
analogues of eq.s\eqn{Uequa},\eqn{fequa}. This investigation is work
in progress \cite{progre} by the author and the same collaborators as in
\cite{solvab1}, \cite{solvab2}.
\par
To illustrate the idea we just recall the results obtained in the
literature for $p$--brane solutions. In \cite{stellebrane} the
following bosonic action was considered:
\begin{eqnarray}
{S^\prime}_{D}&=& \int \, d^Dx \, {\cal L}_D \nonumber \\
 {{\cal L}^\prime}_D   &=& \sqrt{\mbox{det}g} \Bigl (  -2 \, R[g]\, -  \,
  \frac{1}{2} \partial^\mu \phi \, \partial_\mu \phi \nonumber\\
  &&-\frac{(-1)^{n-1}}{2{n!}} \, \exp[-a\phi] \, F_{\mu_1 \dots
  \mu_n}^2 \Bigl )
  \label{paction}
\end{eqnarray}
where $F_{\mu_1 \dots \mu_n}$ is the field strength of an $n-1$--form
gauge potential, $\phi$ is a dilaton and $a$ is some real number. For various values of
$n$ and $a$,  ${S^\prime}_{D}$ is a consistent truncation of some
(maximal or non maximal) supergravity bosonic action $S_D$
 in dimension $D$. By consistent truncation we mean that a
subset of the bosonic fields have been put equal to zero but
in such a way that all solutions of the truncated action are also
solutions of the complete one. The fields that have been
deleted are:
\begin{enumerate}
\item {all the nilpotent scalars}
\item {all the Cartan scalars except that which appears in front of the
$F_{\mu_1 \dots \mu_n}$ kinetic term.}
\item {all the other gauge $q$--form potentials except the chosen
one}
\end{enumerate}
For instance if we choose:
\begin{equation}
a=1 \quad \quad n=\cases{3 \cr 7\cr}
\label{het}
\end{equation}
eq.\eqn{paction} corresponds to the bosonic low energy action of $D=10$
heterotic superstring (N=1, supergravity) where the $E_8\times
E_8$ gauge fields have been deleted. The two choices $3$ or $7$ in
eq.\eqn{het} correspond to the two formulations (electric/magnetic)
of the theory. Other choices correspond to truncations of the type IIA or
type IIB action in the various intermediate dimensions $4\le D\le
10$. Since the $n-1$--form $A_{\mu_1\dots\mu_{n-1}}$ couples to the world volume
of an extended object of dimension:
\begin{equation}
p = n-2
\label{interpre}
\end{equation}
namely a $p$--brane, the choice of the truncated action \eqn{paction}
is motivated by the search for $p$--brane solutions of supergravity.
According with the interpretation \eqn{interpre} we set:
\begin{equation}
\matrix{ n=p+2 & d=p+1 \cr \null &
{\tilde d}= D-p-3     \cr}
\label{wvol}
\end{equation}
where $d$ is the world--volume dimension of the electrically charged
{\it elementary} $p$--brane solution, while ${\tilde d}$ is
the world--volume dimension of a magnetically charged {\it solitonic}
${\tilde p}$--brane with ${\tilde p} = D-p-4$. The distinction between
elementary and solitonic is the following. In the elementary case
the field configuration we shall discuss is a true vacuum solution of
the field equations following from the action \eqn{paction} everywhere in
$D$--dimensional space--time except for a singular locus of dimension
$d$. This locus can be interpreted as the location of an elementary $p$--brane
source that is coupled to supergravity via an electric charge   spread
over its own world volume. In the solitonic case, the field
configuration we shall consider is instead a bona--fide solution of
the supergravity field equations everywhere in space--time without
the need to postulate external elementary sources. The field energy
is however concentrated around a locus of dimension ${\tilde p}$.
Defining:
\begin{equation}
\Delta = a^2 +2\, \frac{d {\tilde d} }{ D-2}
\end{equation}
it was shown in \cite{stellebrane} that action \eqn{paction} admits
the following elementary $p$--brane solution
\begin{eqnarray}
ds^2 & =& \left(1+\frac{k}{ r^{\tilde d} } \right)^{- {4\, { \tilde d}}/{\Delta (D-2)}}
\, dx^\mu \otimes dx^\nu \, \eta_{\mu\nu} \nonumber\\
&& - \left(1+\frac{k}{ r^{\tilde d} } \right)^{  {4\, {d}}/{\Delta (D-2)}}
\, dy^m \otimes dy^n \, \delta_{mn}\nonumber \\
F &= &\lambda (-)^{p+1}\epsilon_{\mu_1\dots\mu_{p+1}} dx^{\mu_1}
\wedge \dots \wedge dx^{\mu_{p+1}} \nonumber\\
&& \wedge \, \frac{y^m \, dy^m}{r} \, \left(1+\frac{k}{r^{\tilde
d}}\right )^{-2} \, \frac{1}{r^{\tilde d}}\nonumber\\
e^{\phi(r)} &=& \left(1+\frac{k}{r^{\tilde d}}\right)^{-2a/\Delta}
\label{elem}
\end{eqnarray}
where $x^\mu$, $(\mu=0,\dots ,p)$ are the coordinates on the $p$--brane world--volume,
$y^m$, $(m=D-d+1,\dots ,D)$ are the transverse coordinates, $r \equiv \sqrt{y^m y_m}$,
$k$ is the value of the electric charge and:
\begin{equation}
\lambda= 2\, \frac{{\tilde d} \, k}{\sqrt{\Delta}}
\end{equation}
The same authors show that that action \eqn{paction} admits also
the following solitonic ${\tilde p}$--brane solution:
\begin{eqnarray}
ds^2 & =& \left(1+\frac{k}{ r^{d} } \right)^{- {4\, {   d}}/{\Delta (D-2)}}
\, dx^\mu \otimes dx^\nu \, \eta_{\mu\nu} \nonumber\\
&& - \left(1+\frac{k}{ r^{ d} } \right)^{  {4\, {\tilde d}}/{\Delta (D-2)}}
\, dy^m \otimes dy^n \, \delta_{mn}\nonumber \\
{\tilde F} &= &\lambda  \epsilon_{\mu_1\dots\mu_{{\tilde d}}p} dx^{\mu_1}
\wedge \dots \wedge dx^{\mu_{\tilde d}}\wedge \frac{y^p}{r^{d+2}} \nonumber\\
e^{\phi(r)} &=& \left(1+\frac{k}{r^{d}}\right)^{2a/\Delta}
\label{solit}
\end{eqnarray}
where the $D-p-2$--form ${\tilde F}$ is the dual of $F$, $k$ is now the magnetic charge
and:
\begin{equation}
\lambda= - 2\, \frac{{\tilde d} \, k}{\sqrt{\Delta}}
\end{equation}
These  $p$--brane configurations are solutions of the second order
field equations obtained by varying the action \eqn{paction}.
However, when \eqn{paction} is the truncation of
a supergravity action both \eqn{elem} and \eqn{solit} are also the
solutions of a {\it first order differential system of equations}.
This happens because they are BPS--extremal $p$--branes which
preserve a fraction of the original supersymmetries. For instance
consider the $10$--dimensional case where:
\begin{equation}
\matrix{D=10 & d=2 & {\tilde d}= 6\cr
a=1 & \Delta = 4 & \lambda = \pm 6 k   \cr}
\label{values}
\end{equation}
so that the elementary string solution reduces to:
\begin{eqnarray}
ds^2 &=& \exp[2 U(r)] \, dx^\mu \otimes dx^\nu - \nonumber\\
&&\exp[-\frac{2 }{3}  U(r)] \, dy^m \otimes dy^m
 \nonumber\\
\exp[2 U(r)]&=& \left(1+\frac{k}{r^6}\right)^{-3/4}\label{stringsol1} \\
F &=& 6k \, \epsilon_{\mu\nu} dx^\mu \wedge dx^\nu \wedge \frac{y^m dy^m}{r} \nonumber\\
&& \times
\left( 1+\frac{k}{r^6}\right)^{-2} \, \frac{1}{r^7}
\label{stringsol2}
\end{eqnarray}
\begin{equation}
\exp[\phi(r)]  =  \left(1+\frac{k}{r^6}\right)^{-1/2}
\label{stringsol3}
\end{equation}
As already pointed out, with the values \eqn{values}, the action \eqn{paction}
is just the truncation of heterotic supergravity where, besides the
fermions, also the $E_8\times E_8$ gauge fields have been set to zero.
In this theory the SUSY rules we have to consider are those of the
gravitino and of the dilatino.
They read:
\begin{eqnarray}
\delta \psi _\mu &=& \nabla_\mu \epsilon\, +\, \frac{1}{96} \,
\exp[\frac{1}{2} \phi] \, \Bigl ( \Gamma_{\lambda\rho\sigma\mu} \nonumber\\
&&+\, 9\,  \Gamma_{\lambda\rho} \, g_{\sigma\mu} \Bigr ) \, F^{\lambda\rho\sigma}
\, \epsilon \nonumber\\
\delta \chi &=& \mbox{i}\, \frac{\sqrt{2}}{4} \, \partial^\mu \phi \,
\Gamma_\mu \epsilon \nonumber\\
&& -\,  \mbox{i}\, \frac{\sqrt{2}}{24} \, \exp[-\frac{1}{2}\phi ] \,
\Gamma_{\mu\nu\rho} \, \epsilon \, F^{\mu\nu\rho}
\label{susvaria}
\end{eqnarray}
Expressing the $10$-dimensional gamma matrices as tensor products of
the $2$--dimensional gamma--matrices $\gamma_\mu$ ($\mu=0,1$) on the
$1$--brane world sheet with the $8$--dimensional gamma--matrices
$\Sigma_m$ ($m=2,\dots, 9$) on the transverse space it is easy to check
that in the background \eqn{stringsol1}, \eqn{stringsol2},\eqn{stringsol3}
the SUSY variations
\eqn{susvaria} vanish for the following choice of the parameter:
\begin{equation}
\epsilon   =   \left( 1+\frac{k}{r^6}\right)^{-3/16} \, \epsilon_0
\otimes \eta_0
\label{carmenpara}
\end{equation}
where the constant spinors $\epsilon_0$ and $\eta_0$ are respectively
$2$--component and $16$--component and have both positive chirality:
\begin{equation}
\matrix{ \gamma_3 \, \epsilon_0 = \epsilon_0 &  \Sigma_0 \, \eta_0 = \eta_0}
\label{chiralcondo}
\end{equation}
Eq.\eqn{chiralcondo} is the $D=10$ analogue of eq.\eqn{parcondicio}.
Hence we conclude that the extremal $p$--brane solutions of all
maximal (and non maximal) supergravities can be
obtained by imposing the supersymmetry invariance of the background
with respect to a projected SUSY parameter of the type
\eqn{carmenpara}. \par
In the maximal case a general analysis of the resulting evolution equation
for the scalar fields in the solvable Lie algebra representation is
work in progress \cite{progre}.



\begin{thebibliography}{99}
\bibitem{sumschwarz} For a summary of the vast recent literature on
these developments started by E. Witten  and several other authors
we refer the reader to the lectures
by J.H. Schwarz (hep-th/9607201) and also to all the other lectures
at the Trieste Spring School 1996. (Nucl. Phys. Proc. Suppl. to appear).
\bibitem{mylec}For a review see: P. Fr\'e, Lectures on Special K\"ahler Geometry
and Electric--Magnetic Duality Rotations,   Nucl. Phys. B (Proc.
Suppl.) 45B,C (1996) 59-114
\bibitem{BPS} R. Dijkgraaf, E. Verlinde, H. Verlinde, hep-th 9603126,
A. Strominger, C. Vafa, hep-th 9601029,
R.R. Khuri, hep-th 9609094.
\bibitem{topftwist_1}
D.~Anselmi and P.~Fr\'e, Nucl. Phys. B392 (1993) 401.
\bibitem{topftwist_2}
D.~Anselmi and P.~Fr\'e, Nucl. Phys. B404 (1993) 288;
Nucl. Phys. B416 (1994) 255.
\bibitem{topf4d_8} D.~Anselmi and P.~Fr\'e, Phys. Lett. B347 (1995) 247.
\bibitem{wittft} E. Witten, Comm. Math. Phys. {\bf 117} (1988) 353,
 E. Witten, Comm. Math. Phys. {\bf 118} (1988) 411.
\bibitem{jgpnoi} L. Andrianopoli, M. Bertolini, A. Ceresole, R.
D'Auria, S. Ferrara, P. Fr\'e, hep-th 9603004 to appear on
Nucl. Phys. B  and L. Andrianopoli, M. Bertolini, A. Ceresole, R.
D'Auria, S. Ferrara, P. Fr\'e, T. Magri,  hep-th/9605032,
to appear on Journal of Geometry and Physics.
\bibitem{specspec2}
B. de Wit, P. G. Lauwers and A. Van Proeyen Nucl. Phys. {\bf B255}
(1985) 569.
\bibitem{skgsugra_4}
L.~Castellani, R.~D'Auria and S.~Ferrara,
Phys. Lett. 241B (1990) 57;
Class. Quantum Grav. 7 (1990) 1767.
\bibitem{skgsugra_1}
R.~D'Auria, S.~Ferrara and P.~Fr\'e,
Nucl. Phys. B359 (1991) 705.
\bibitem{cardoso} G. L. Cardoso, D. Luest, T. Mohaupt, hep-th/9608099
\bibitem{kalvanp1} R. Kallosh, A. Linde, T. Ortin, A. Peet, A. Van
Proeyen, Phys. Rev. D46, (1992) 5278
\bibitem{ferkal2} S. Ferrara, R. Kallosh, A. Strominger, hep-th/ 9508072
\bibitem{strom3} A. Strominger hep-th/9602111
\bibitem{ferkal4} S. Ferrara, R. Kallosh, hep-th/9602136
\bibitem{kalmany} K. Berhrdt, R. Kallosh, J. Rahmfeld, M. Shmakova,
Win Kai Wong, hep-th/9608059
\bibitem{carbon} R. Kallosh, B. Kol hep th 9602014
\bibitem{cereferpro} A. Ceresole, R. D'Auria, S. Ferrara, Proceedings
of the Trieste workshop on Mirror Symmetry and S--Duality, Trieste 1995,
Ed.s Narain and E. Gava. hep--th 9509160
\bibitem{noi} M. Bill\'o, R. D'Auria, S. Ferrara, P. Fr\`e, P. Soriani
and A. Van Proeyen,  R-symmetry and the topological twist of N=2
effective supergravities of heterotic strings, hep-th 9505123, to
appear on IJMP.
\bibitem{witmono} E. Witten Math. Res. Lett. {\bf 1} (1994) 484.
\bibitem{seiwit} N. Seiberg, E. Witten, Nucl. Phys. {\bf B426} (1994) 19 and
Nucl. Phys. {\bf B431} (1994) 484
\bibitem{solvab1} L. Andrianopoli, R. D'Auria, S. Ferrara, P. Fr\'e and M. Trigiante,
{\it R-R Scalars, U-Duality and
Solvable Lie Algebras} , hep-th/9611014
\bibitem{solvab2} L. Andrianopoli, R. D'Auria, S. Ferrara, P. Fr\'e, R. Minasian
and M. Trigiante,
 {\it Solvable Lie Algebras
in Type IIA, Type IIB and M Theories} , hep-th 9612202
\bibitem{sase}
A. Salam and E. Sezgin,  {\it Supergravities in diverse Dimensions}
Edited by A. Salam and E. Sezgin, North--Holland, World Scientific
1989, vol. 1
\bibitem{cre} E. Cremmer, in  Supergravity '81,
ed. by S. Ferrara and J.G. Taylor, pag. 313
\bibitem{huto} { C.M. Hull and P.K. Townsend,  Nucl. Phys. {\bf B438} (1995) 109.}
\bibitem{helgason} S. Helgason,   Differential Geometry and Symmetric Spaces,
New York: Academic Press (1962).
\bibitem{alex} D.V. Alekseevskii, Math. Izv. USSR, {\bf Vol. 9} (1975), No.2
\bibitem{voialtri}L. Andrianopoli, R. D'Auria and S. Ferrara,
{\it U--Duality and Central Charges
in Various Dimensions Revisited}, hep-th 9612105
\bibitem{stellebrane} H.Lu, C.N. Pope, E. Sezgin and K.S. Stelle, hep
th 9508042, H. Lu, C.N. Pope, hep th 9605082, M. J. Duff, H. Lu and
C.N. Pope, hep th 9604052, H.Lu, C.N. Pope and K.S. Stelle hep the
9602140.
\bibitem{progre}L. Andrianopoli, R. D'Auria, S. Ferrara, P. Fr\'e and M. Trigiante,
work in progress.
\end{thebibliography}
\end{document}